\documentclass[twocolumn,resetfootnote]{aastex701}

\newcommand{\hi}{H\,{\footnotesize I}}

\newcommand{\disperse}{\textsc{DisPerSE}}
\newcommand{\rvir}{R_{\rm 200c}}
\newcommand{\msunh}{\mathrm{M_\odot}h^{-1}}
\newcommand{\mpch}{\mathrm{Mpc}~h^{-1}}
\newcommand{\kpch}{\mathrm{kpc}~h^{-1}}

\begin{document}

\title{Detection of \hi~filament: Pair Stacking  vs. Filament Stacking}

\newcommand{\NAOC}{National Astronomical Observatories, Chinese Academy of Sciences, Beijing 100101, China}
\newcommand{\UCAS}{School of Astronomy and Space Science, University of Chinese Academy of Sciences, Beijing 100049, China}

\author[0009-0004-6895-6743]{Yuxi Meng}
\affiliation{\NAOC}
\affiliation{\UCAS}
\email[show]{yxmeng@bao.ac.cn}

\author[0000-0002-9937-2351]{Jie Wang}
\affiliation{\NAOC}
\affiliation{\UCAS}
\email[show]{jie.wang@nao.cas.cn}

\author[0000-0003-3433-8416]{Yingjie Jing}
\affiliation{\NAOC}
\affiliation{\UCAS}
\email[]{}

\author[0009-0001-2951-7383]{Hongxiang Chen}
\affiliation{\NAOC}
\affiliation{\UCAS}
\email[]{}

\author[0009-0004-6617-6903]{Zerui Liu}
\affiliation{\NAOC}
\affiliation{\UCAS}
\email[]{liuzr@bao.ac.cn}

\begin{abstract}

The faint 21 cm signal emitted by neutral hydrogen in cosmic filaments is expected to be detectable. However, due to its weakness, stacking techniques are required. We assessed two stacking methods—pair stacking and filament stacking—using the EAGLE and IllustrisTNG simulations. Pair stacking leverages the fact that cosmic filaments connect massive structures (i.e., knots) in the cosmic web, while filament stacking directly aggregates filaments identified from galaxy distributions. Our analysis indicates that, although pair stacking is convenient, it faces contamination from massive structures; after removing this contamination, the filament signal is significantly reduced. In contrast, \hi~detection via filament stacking appears more promising. The column density in filament stacking reaches $\sim 10^{16}$--$10^{17}~\mathrm{cm}^{-2}$ even when all haloes are masked, whereas pair stacking does not reach this level even without masking, and is further suppressed by several orders of magnitude once masking is applied. The effectiveness of filament stacking can be further improved with higher galaxy number density and better spatial resolution in radio intensity mapping observations. With the advent of upcoming optical and radio data, the detection of \hi~in cosmic filaments remains promising.

\end{abstract}



\section{Introduction}

Neutral hydrogen (\hi), the most prevalent element in the universe, is crucial for galaxy star formation and serves as an effective tracer of baryon distribution. Cosmological simulations suggest the majority of \hi~is found within galaxies and the circum-galactic medium (CGM), with smaller amounts existing as clouds or clumps in the intergalactic medium (IGM). Large-scale cosmic filaments primarily consist of the photoionized warm-hot intergalactic medium (WHIM). Although \hi~is not the main component of the IGM, its presence is confirmed by Ly$\alpha$ absorption studies, indicating that \hi~may significantly influence star formation in galaxies. Specifically, the \hi~clouds located in cosmic filaments can move towards clusters, promoting rapid star formation in galaxies via cold accretion at high redshifts.

Detecting \hi~in cosmic filaments is challenging because of its low density. In recent years, several 21 cm intensity mapping experiments and surveys \citep[][]{2012IJMPS..12..256C, 2015aska.confE..19S, 2016mks..confE..32S, 2023ApJ...954..139L} have opened a new observational window for probing large-scale \hi~without resolving individual galaxies. While these surveys are designed to measure large-scale \hi~fluctuations, isolating the faint filamentary signal remains particularly difficult. One promising strategy to enhance detectability is stacking, which improves the signal-to-noise ratio (S/N) by combining measurements from many systems. Filament detection through the stacking of galaxy pairs is straightforward to implement and relatively insensitive to the choice of pairs. This technique has been employed in the detection of dark matter \citep[][]{2016MNRAS.457.2391C}, magnetic fields \citep[][]{2021MNRAS.505.4178V}, mass-to-light ratio \citep[][]{2020MNRAS.498.3158Y}, WHIM \citep[][]{2019A&A...624A..48D}, Ly$\alpha$ emission \citep[][]{2018MNRAS.475.3854G}, and shock fronts \citep[][]{2023SciA....9E7233V}. However, using pair stacking for \hi~detection \citep[][]{2019MNRAS.489..385T} has not yielded high signals due to the extremely low \hi~column density in cosmic filaments. Recent hydrodynamical simulations based on IllustrisTNG, combined with mock FAST observations, predict a beam-smoothed filament signal of only $\sim0.3\,\mu$K after background subtraction \citep[][]{2025ApJ...984..177L}, highlighting the severe observational challenge for this technique. Thus, a more effective stacking technique is necessary for \hi~detection in cosmic filament.

The effectiveness of a stacking technique, as determined by the S/N, largely hinges on the precision of stacking genuine cosmic filaments. Identifying filaments using galaxy distributions holds the potential for greater accuracy than merely pairing galaxies. Such stacking methods have been leveraged in discovering dark matter and the WHIM \citep[][]{2020A&A...637A..41T, 2020A&A...643L...2T, 2022A&A...667A.161T}. Yet, deriving accurate filament identifications from galaxy distributions can suffer due to incomplete galaxy catalogs. Various filament finders employing diverse algorithms have been developed, but their effectiveness is often debated. Consequently, we conduct a test to assess the effectiveness of both pair stacking and filament stacking.

The latest cosmological hydrodynamical simulations are valuable for evaluating stacking techniques. They not only deliver galaxy populations with accurate properties but also represent the baryonic content in the CGM and IGM by integrating key physical processes affecting these regions. Consequently, they predict the \hi~content we focus on. The scale of these simulations is extensive, covering sufficient cosmic filaments spanning tens of megaparsecs, enabling more reliable predictions of filament signals.

This paper is organized as follows: Section \ref{sec:methods} provides a concise introduction to the simulations utilized and outlines the stacking methodology. In Section \ref{sec:results}, we explore the elements influencing the effectiveness of the stacking techniques and evaluate their capabilities. The discussion of results is presented in Section \ref{sec:discussions}, followed by conclusions in Section \ref{sec:conclusions}.

\section{Methods}
\label{sec:methods}

\subsection{Simulations}
We utilize the EAGLE (Evolution and Assembly of GaLaxies and their Environments) simulations alongside the IllustrisTNG (The Next Generation Illustris Simulations) to conduct our analysis. Each set employs distinct simulation algorithms and subgrid physics. This allows us to gauge the uncertainty stemming from varying galaxy formation models. Our focus on \hi~in the local universe prompts us to use the z=0 snapshots from both simulations.

The EAGLE simulations \citep[][]{2015MNRAS.446..521S, 2015MNRAS.450.1937C, 2016A&C....15...72M} are a series of hydrodynamical simulations executed with a significantly enhanced version of the tree-SPH code GADGET-3. This simulation suite includes subgrid physics such as radiative cooling, photoheating, star formation, stellar evolution, metal enrichment, stellar feedback, black hole growth, and AGN feedback. It's calibrated against the stellar mass function, galaxy sizes, and the black hole–stellar mass relation of z=0 galaxies. In the EAGLE simulations, groups and subhalos are identified using the friends-of-friends (FOF) algorithm and the SUBFIND algorithm, respectively. We conducted postprocessing to derive the \hi~mass for each gas particle in accordance with \citet{2013MNRAS.430.2427R}. EAGLE employs cosmological parameters derived from Planck 2014 ($\Omega_{m,0}=0.307$, $\Omega_{\Lambda,0}=0.693$, $\Omega_{b,0}=0.04825$, $h=0.6777$, $\sigma_8=0.8288$, $n_s=0.9611$). We utilized the reference simulation REFL0100N1504, which features a 100 comoving Mpc box on each side, and contains $1504^3$ dark matter particles with masses of $9.7\times 10^6 \mathrm{M_\odot}$ and $1504^3$ gas particles with masses of $1.81\times 10^6 \mathrm{M_\odot}$.

\begin{table}
\caption{Volume fraction ($f_\mathrm{V}$) and \hi~mass fraction ($f_\mathrm{HI}$) in knots, filaments, and voids for the EAGLE and TNG100 simulations. Note that the definition of the filament boundary affects the inferred $f_\mathrm{HI}$: adopting filament radii in the range $0.5$-$2~\mathrm{Mpc}/h$ results in $f_\mathrm{HI}$ values of $\sim36$-$65\%$.}
\label{tab:fraction}
\centering
\begin{tabular}{lcccc}
\hline
Simulation & Environment & $f_\mathrm{V}$ (\%) & $f_\mathrm{HI}$ (\%) \\
\hline
EAGLE   & Knots     & 2.34  & 21.59 \\
        & Filaments & 6.35  & 53.41 \\
        & Voids     & 91.31 & 25.00 \\
\hline
TNG100  & Knots     & 2.18  & 18.05 \\
        & Filaments & 6.13  & 47.79 \\
        & Voids     & 91.69 & 34.16 \\
\hline
\end{tabular}
\end{table}

The IllustrisTNG simulations \citep[][]{2019ComAC...6....2N} utilize the moving-mesh code AREPO. Compared to the original Illustris simulation, the galaxy formation model has been improved, particularly in kinetic AGN feedback, galactic winds, and the incorporation of magnetic fields. Groups and subhalos are detected using the FOF and SUBFIND algorithms. Neutral hydrogen abundance is calculated on-the-fly as per \citet{2013MNRAS.430.2427R}, and we subsequently conducted post-processing to determine the \hi~mass of the gas cells. The IllustrisTNG simulation adopts cosmological parameters based on Planck 2015 data ($\Omega_{m,0}=0.3089$, $\Omega_{\Lambda, 0}=0.6911$, $\Omega_{b,0}=0.0486$, $h=0.6774$, $\sigma_8=0.8159$, $n_s=0.9667$), which show slight deviations from Planck 2014. We applied the highest resolution run, TNG100-1, within the TNG100 series, encompassing a box size of 106.5 comoving Mpc per side, containing $1820^3$ dark matter particles of $7.5\times 10^6~\mathrm{M_\odot}$, alongside $1820^3$ gas cells of $1.4\times 10^6~\mathrm{M_\odot}$.

\begin{figure*}
    \centering
    \includegraphics[width=\textwidth]{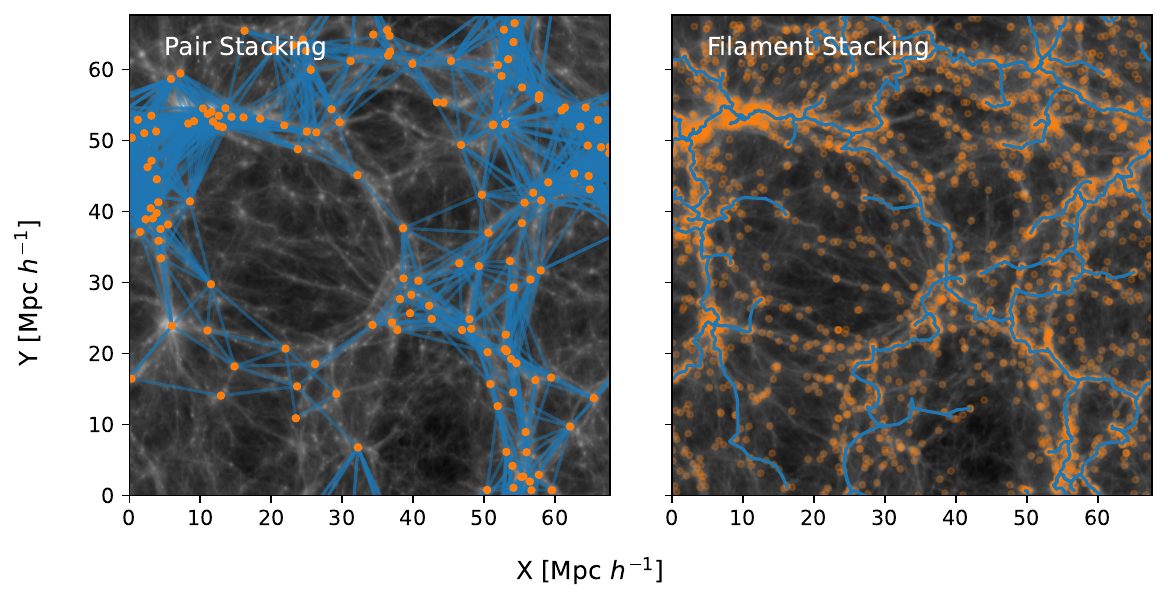}
    \caption{A demonstration of galaxy pairs (left panel) and filaments identified with galaxy distribution (right panel) in EAGLE simulation. Orange points represent galaxies in both panels, and blue lines represent galaxy pairs (left panel) or cosmic filaments (right panel). For clarity, the galaxy pairs are selected from the most massive galaxies of  $M_* \geq 10^{11}~\msunh$. The filaments are identified with galaxies with  $M_* \geq 10^{10}~\msunh$ using \disperse.}
    \label{fig:illustration}
\end{figure*}

In our simulation, we classify the space into different types called knots, filaments, and voids. The knots are defined as areas within $5\times \rvir$ around halos with masses $M_h \geq 10^{13}~\msunh$, where $\rvir$ is the radius enclosing a mean density 200 times the cosmic critical density. Filaments are designated as zones within $1~\mpch$ surrounding filaments identified by galaxies with stellar masses $M_* \geq 10^8~\msunh$ using \disperse, while voids encompass the remaining portions. We observed that both the volume fraction and the \hi~mass fraction align consistently between the EAGLE and TNG simulations (see Table~\ref{tab:fraction}). Knots, being the densest structures, comprise $\sim20\%$ of the \hi~content despite occupying just $\sim2\%$ of the volume. Filaments occupy $\sim7\%$ of the total volume and account for $\sim50\%$ of the \hi. Meanwhile, the voids, which cover $\sim91\%$ of the total volume, contain only $\sim30\%$ of the \hi.

\subsection{Stacking Techniques}

\subsubsection{Pair Stacking}

The pair stacking technique is developed to enhance the signal emitted from the filaments which are identified by the straight lines between a pair of galaxies \citep[][]{2016MNRAS.457.2391C}. It is based on the assumption that galaxies residing in the most massive halos are preferentially connected by cosmic filaments. Therefore, by selecting pairs of galaxies with proper separations, the connection between pairs could be identified as the filaments in the cosmic webs (see left panel of Figure \ref{fig:illustration}).

Additional \hi~sources in the vicinity of galaxy pairs may coincidentally align with the filaments axis (i.e., the straight line between the pair), thereby contaminating the filament signal. Assuming these structures are approximately spherically distributed around the ends of the filaments in the stacked map, their contribution can be subtracted from the total flux, resulting in a signal predominantly from cosmic filaments between the pairs, which peaks centrally in the stacked map.

To evaluate the detectability of filaments with pair stacking, we perform the stacking procedure following observational analyses. The procedures of stacking are as follows:

1. \textit{Create a catalogue of galaxy pairs for the purpose of stacking.} The catalogue is generated based on the galaxy catalogue and specific criteria for selecting pairs.
Specifically, central galaxies with stellar mass $M_* \geq 10^{10}~\msunh$ are selected, mimicking magnitude-limited samples in observational studies. Two galaxies are considered a pair if their projected separation lies between $6\text{--}14~\mpch$, and their line-of-sight distance is less than $6~\mpch$ , following, e.g., \citet{2019MNRAS.489..385T, 2019A&A...624A..48D}.

2. \textit{Create maps of column density for each galaxy pair separately.} To create a column density map, we first construct a cube of \hi~density on a $0.1~\mpch$ grid. 
For each galaxy pair, we extract a box-shaped region around the pair, with a thickness of $20~\mpch$ along the line of sight. 
The extracted box is then projected along the line of sight to produce the column density map. The map is subsequently rotated such that the line connecting the two galaxies is aligned with the horizontal axis. Finally, a fixed subregion is extracted from the rotated map, serving as a rescaling step that ensures the two galaxies occupy the same relative positions across all maps.

3. \textit{Stack} column density maps and evaluate the spherical profile around the ends.
We stack maps for each galaxy pair to obtain a column density map around the filaments, which includes the contribution from both the filaments and nearby contaminating galaxies. By assembling these maps and averaging each with its mirrored version, we produce a symmetrical map (hereafter referred to as the stacked map). From this stacked map, we extract the column density values and the distances to each galaxy in the pair from pixels within sectors perpendicular to the line connecting the pair. 
Each sector is centered on one of the two galaxies (i.e., the vertex coincides with the galaxy position) and spans a $\pm 45^\circ$ opening angle around the perpendicular direction. We then perform a least-squares fit to the column density values in these sectors to determine the spherical profile $\sigma(r)$, representing the contamination contribution. The fitting function is a third-order (cubic) polynomial modulated by an exponential factor.

4. \textit{Extract} the filament signal and estimate its uncertainties. we subtract the fitted profile from the stacked map and obtains a residual map. The residual map reveals a filament-like structure bridging both ends of the pair, with its peak intensity situated at the map's center. We select a square region centered on the pair, with a side length equal to one-fifth of the pair separation on the residual map. The mean column density within this region is taken as the filament signal.
\footnote{We have verified that the extracted filament signal is not sensitive to the size of the square region, as the central filament signal in the residual map is relatively smoothly distributed.}
We apply these steps across $X$, $Y$, and $Z$ projection directions. To estimate uncertainties, we divide the simulation box into subvolumes and perform jackknife resampling by excluding, in turn, pairs whose centers fall within each subvolume.

Following our procedures, we determined the column density of the filament signal along with an estimation of its uncertainties. In observations, the measured flux includes contributions from the filament signals, foreground contamination, and instrumental noise. Nonetheless, the filament signal strength measured reflects the intrinsic signal achievable through the pair stacking technique. A stronger filament signal implies a higher likelihood of detection using this technique.

\subsubsection{Filament Stacking}

Filament stacking utilizes the pattern inherent in the distribution of galaxies themselves, rather than tracing cosmic filaments with pairs of galaxies. Various algorithms are designed to detect cosmic filaments within galaxy survey catalogues. Among these, \disperse~\citep[][]{2011MNRAS.414..350S, 2011MNRAS.414..384S} excels due to its foundation in discrete Morse theory and persistence theory, enabling reliable identification of topological structures, such as cosmic filaments, from galaxy distributions (see right panel of Figure \ref{fig:illustration}).

\disperse~begins by estimating a density field from the distribution of discrete tracer particles like galaxies and then identifies critical points—maxima, saddle points, and minima—while tracing ridge-like integral lines between maxima and saddle points, which serve as filament spines. Each filament is assigned a persistence ratio, defined as the density ratio between the connected maximum and saddle point, which represents the robustness of the filament. The ratio is expressed in units of $\sigma$, defined analogously to a Gaussian random field, with higher values indicating more robust filaments. Filaments with persistence ratio under a given threshold, expressed in number of $\sigma$s, are filtered out to retain only the strongest filaments.

Based on the identified filaments, the filament stacking technique is to stack the pixels (or voxels in 3D) directly according to their distance from the filament spines. Specifically, we apply filament stacking to the simulation with the following prescriptions:

1. \textit{Generate column density and galaxy maps from simulation.} We take a slice that is $20~\mpch$ thick from the simulation box, matching the thickness used in our pair stacking method. Then we project and create a two-dimensional distribution of galaxies and a \hi~column density map. Only galaxies above a specific stellar mass cutoff (e.g., $M_* \geq 10^{8}$, $10^{9}$, or $10^{10}~\msunh$, corresponding to galaxy number densities of $n_{\rm gal}= 1.1\times10^{-1}$, $3.5\times10^{-2}$, and $8.8\times10^{-3}~\mathrm{Mpc}^{-3}h^{3}$ in the EAGLE simulation) are included, and the column density map is divided into pixels measuring $0.01~\mpch$ each. \footnote{To better resolve the stacked filament structure, we adopt a higher spatial resolution here. We have verified that the results converge to those obtained with lower resolution at larger radii.}

2. \textit{Identify the cosmic filaments with \disperse.}
To ontain a filament catalogue with \disperse, the procedure relies on two main inputs: the tracer distribution and the persistence threshold. In our research, we use the 2D galaxy distribution generated in the previous step as tracers. We select filaments that have persistence ratios greater than $3\sigma$, to filter out spurious structures.
Figure~\ref{fig:illustration} illustrates the tracers as orange dots and the detected filament spines as blue curves.

3. \textit{Extract the filament column density profile and evaluate its uncertainties.}
Based on the column density map and filament catalogue, we sort pixels in the column density map by their distance to the filament spine, then get the radial profile of \hi~column density perpendicular to the filament spine. We obtain profiles from multiple 2D slices taken along the $X$, $Y$, and $Z$ directions, and compute their average. The uncertainty is estimated using the jackknife resampling method.

After this procedure, we obtain the column density profile, which demonstrates the capability of the filament stacking technique. Since the column density is measured in the same units as the filament signal obtained via pair stacking, we can evaluate the effectiveness of both stacking approaches by comparing their predicted column density values directly.

\subsection{Mask halo contribution}
\label{subsec:mask}

\hi~in the universe resides in different structures, including the clumps in the IGM and the massive galaxies. \hi~in galaxies has column densities as high as $\sim 10^{20}~\mathrm{cm^{-2}}$, while \hi~clumps in the IGM are only $\sim 10^{15}~\mathrm{cm^{-2}}$, orders of magnitude lower than in galaxies. Therefore, it is necessary to evaluate the contribution of \hi~in different structures to the stacked signal.

A way to analyze the \hi~component in the stacked signal is to mask out haloes above a given mass threshold, together with their surrounding regions \citep[e.g.,][]{2022A&A...667A.161T}. We find that $\sim10\%$ of \hi~resides within $2\times\rvir$ of group-size haloes with $M_h \geq 10^{13}~\msunh$, which occupy only $\sim0.16\%$ of the total volume. Meanwhile, haloes more massive than MW-size systems, i.e., $M_h \geq 10^{12}~\msunh$, contain $\sim39\%$ of the total \hi~while occupying $\sim0.26\%$ of the volume. In addition, $\sim80\%$ of \hi~resides in haloes with $M_h \geq 10^{11}~\msunh$, while these regions account for only $\sim0.33\%$ of the volume.

For both pair and filament stacking, we mask pixels within these halo regions, excluding them from the column density maps and from the computation of the mean column density in both stacking methods. The goal is to estimate the mean \hi~density outside the vicinity of haloes. We further assess the effectiveness of the two stacking methods with and without masking.

Note that masking haloes differs from masking only massive galaxies, since it inevitably removes satellite galaxies (some of which are dwarf galaxies below the detection limit) and the \hi~clumps near the haloes. We also performed galaxy masking according to their stellar masses; the results are presented in the Appendix \ref{app:mask_gal}. In \hi~observations with limited spatial resolution, bright galaxies and dwarfs are difficult to separate, leading to a so-called confusion effect, and in this context halo masking is more practical. In the main results of our study, we only analyze the contribution of haloes instead of individual galaxies.

\section{Results}
\label{sec:results}

\subsection{Pair Stacking}
By stacking pairs selected from the galaxy catalogue, we obtain an average column density map, in which a high-density bridge connects two peaks of density at the ends of the pair. The central region between the pair has a total column density of $1.81\times 10^{17}~\mathrm{cm^{-2}}$, including a dominant $\sim75\%$ ($1.35\times 10^{17}~\mathrm{cm^{-2}}$) contribution from background, estimated from a square region of the same size as the signal region but located far from the galaxy pair on the stacked map, without subtracting the profile component, and an additional $\sim15\%$ ($1.62\times 10^{17}~\mathrm{cm^{-2}}$) contribution from the spherical profile around both ends of the pair. The remaining $\sim10\%$ ($\sim 1.98\times 10^{16} \mathrm{cm^{-2}}$) is expected to represent cosmic filaments.

Upon examining the column density maps of individual galaxy pairs, we find that the stacked map is primarily contributed by three types of pairs. First, there are pairs of galaxies connected by cosmic filaments, as anticipated. Stacking these pairs enables a statistical estimate of the average \hi~column density within cosmic filaments. Second, we find pairs where only a void exists between the galaxies. The absence of filaments in such pairs highlights the limitations of using galaxy pairs to trace cosmic filaments. These void-related pairs are not associated with filaments and exhibit low \hi~column densities, such that their inclusion can reduce the S/N. Third, there are cases where a high-density structure, such as a massive halo, is located between the galaxies. Although these cases are not dominant, they significantly contaminate the stacked signal because their \hi~column density is considerably higher than that found in the IGM. While subtracting the spherical profile can partially mitigate this contribution, a significant residual contamination remains.

\begin{figure}
    \centering
    \includegraphics[width=\columnwidth]{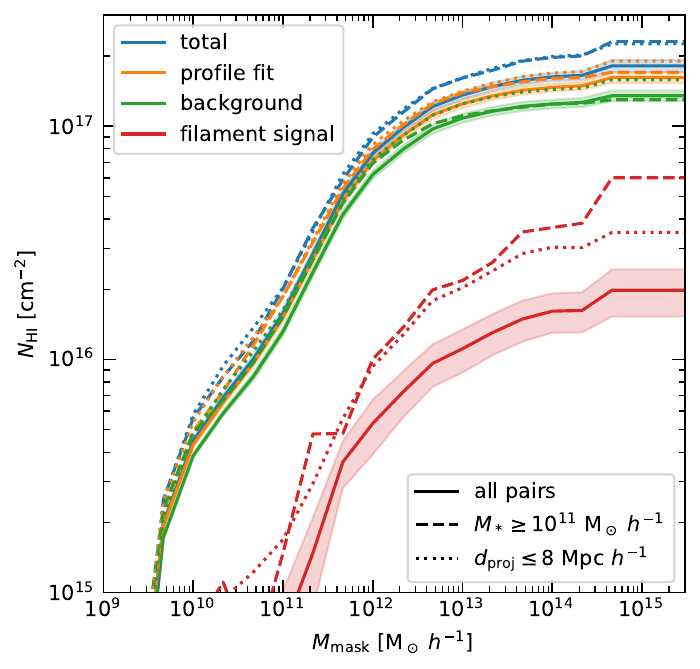}
    \caption{Different components of the stacked column density obtained from pair stacking} and their dependence on the halo mass threshold used for masking. The blue, orange, green and red curves denote total density, fitted spherical profile, background density level, and filament signal, respectively. Solid curves correspond to all pairs, dashed curves represents massive pairs with $ M_* \geq 10^{11}~\msunh$, and dotted curves are obtained from a subsample of pairs with separations shorter than $8~\mpch$. 
    \label{fig:Pstack}
\end{figure}

We first estimate the contamination from massive haloes using a masking procedure (see Section~\ref{subsec:mask}). Figure~\ref{fig:Pstack} shows the relationship between column density and the lower halo mass limit used for masking. The solid curves represent the column density stacked across all pairs, while the blue and red curves denote the total column density and the residual filament signal, respectively. The impact of halos with masses $M_h \geq 10^{15}~\msunh$ on the stacking results is negligible, due to their low number density. A noticeable decrease in the stacked column density is observed when progressively masking lower-mass haloes, reflecting their higher abundance. Excluding cluster-size halos with masses $M_h \geq 10^{14}~\msunh$ reduces the stacked column density; the average total column density decreases to $90\%$ ($1.63\times 10^{17}~\mathrm{cm^{-2}}$), while the filament signal decreases to $81\%$ ($1.61\times 10^{16}~\mathrm{cm^{-2}}$). Masking halos within the mass range of $M_h = 10^{11\text{--}14}~\msunh$ further reduces the total column density to $9\%$ ($1.58\times 10^{16}~\mathrm{cm^{-2}}$). When group-size halos with $M_h \geq 10^{13}~\msunh$ are masked, the filament signal decreases to $56\%$ ($1.11\times10^{16} \mathrm{cm^{-2}}$) of the unmasked level. Further masking of MW-size halos with $M_h \geq 10^{12}~\msunh$ lowers the filament signal to $27\%$ ($5.31\times 10^{15}~\mathrm{cm^{-2}}$). Finally, the filament signal drops below $10^{15}~\mathrm{cm^{-2}}$ ($\sim5\%$) when excluding halos with $M_h \geq 10^{11}~\msunh$. Overall, these results indicate that detecting the IGM \hi~signal within filaments in the stacked analyses of all pair samples is highly challenging due to substantial halo contamination.

Regarding the other issue of S/N degradation associated with void-related pairs, we attempt to reduce their fraction by carefully selecting subsets of galaxy pairs, thereby enhance the stacked filament signal. However, it is also important to control for halo contamination, as preferentially selecting pairs in dense environments can increase the contribution from massive haloes. Moreover, selecting too few pairs may increase statistical noise in actual observations. If the selection does not significantly boost the intrinsic filament signal, the net effect may still be a degradation of the S/N. Therefore, it is essential to assess whether the filament signal can be sufficiently enhanced when selecting specific subsets of galaxy pairs.

In Figure \ref{fig:Pstack}, we illustrate two distinct pair selection strategies, in which are described as follows. The first strategy involves selecting pairs based on the stellar mass of galaxies at each end. Generally, more massive pairs are found in denser regions and may align more closely with cosmic filaments. Pairs with galaxy $M_* \geq 3\times 10^{10}~\msunh$ exhibit a total column density that is $5\%$ greater than that of all pairs, though the filament signal column density shows only a slight difference. For the most massive pairs, with galaxies of $M_* \geq 10^{11}~\msunh$, shown by the dashed curves in Figure \ref{fig:Pstack}, the total column density increases by $28\%$ ($2.31\times 10^{17}~\mathrm{cm^{-2}}$) without masking, and their filament signal density is $6.01\times 10^{16}~\mathrm{cm^{-2}}$, approximately three times that of all pairs. However, most of this apparent enhancement in filament signal density is due to \hi~in massive halos, as the difference is significantly reduced when halos of $M_h \geq 10^{12}~\msunh$ are masked. 

The second strategy focuses on pair separation, indicating a different filament population. Short pairs tends to trace shorter but more robust filaments in denser environments. By stacking pairs with separations less than $8~\mpch$ (dotted curves in Figure \ref{fig:Pstack}), we achieve an total column density that is comparable to the most massive pairs, with a filament signal density of $3.50\times 10^{16}~\mathrm{cm^{-2}}$, $75\%$ higher than that of all pairs in the unmasked case. However, this apparent enhancement in the filament signal becomes much less significant when masking halos of $M_h \geq 10^{12}~\msunh$, suggesting that the increase is primarily driven by halo contamination. Longer pairs show column densities that are comparable to or lower than those of all pairs.

In summary, by carefully choosing the subsamples, the unmasked signal signal can be significantly enhanced; however, separating it from halo contamination remains challenging.

\begin{figure}
    \centering
    \includegraphics[width=\columnwidth]{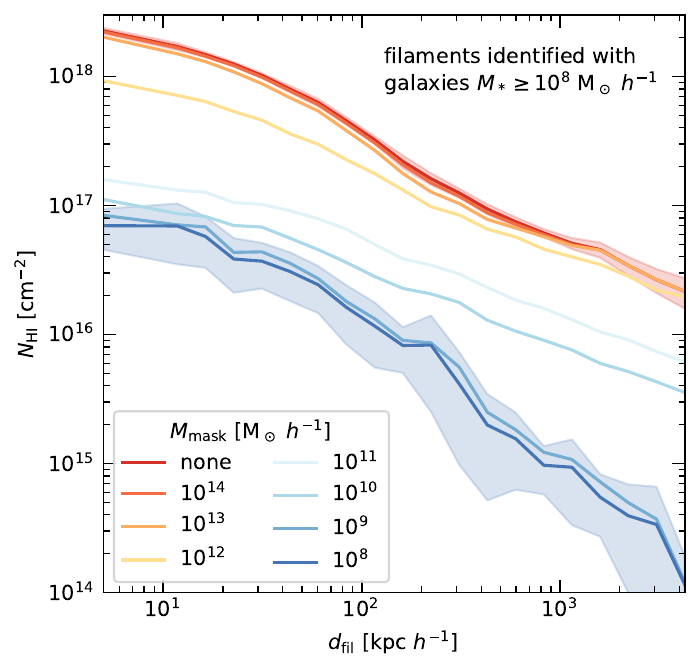}
    \caption{Density profile obtained with filament stacking, shown for different halo masking strengths in EAGLE simulation. Filaments are identified with galaxies of $ M_* \geq 10^{8}~\msunh$, and curves from red to blue represent increasing masking strength, from no masking to masking $2\times \rvir$ around halos with $ M_h \geq 10^{8}~\msunh$. Note that the innermost point corresponds to the center of the first radial bin ($0$--$0.01~\mpch$). The map pixel size is  $0.01~\mpch$, therefore this does not imply resolution below the pixel scale.}
    \label{fig:Fstack_mask}
\end{figure}

\begin{figure}
    \centering
    \includegraphics[width=\columnwidth]{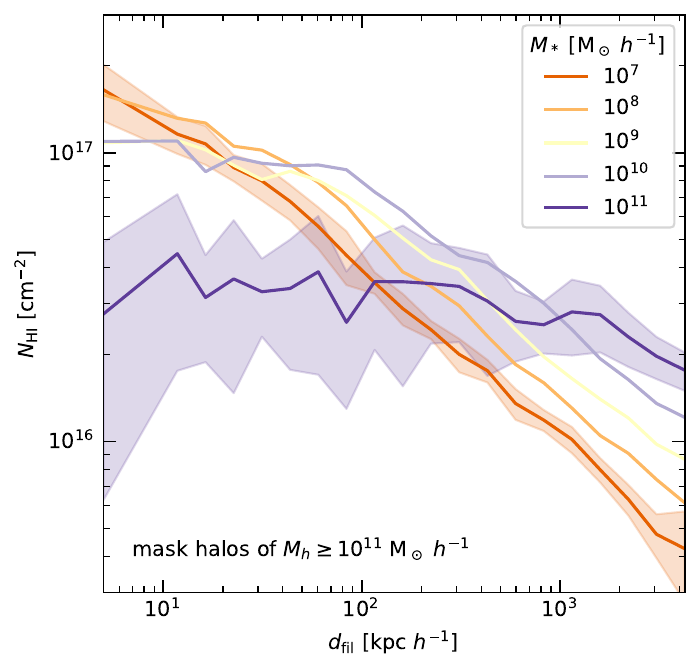}
    \caption{Density profile obtained with filament stacking, shown for different galaxy number densities (corresponding to varying stellar mass thresholds) in EAGLE simulation. Regions within $2\times \rvir$ around halos of $ M_h \geq 10^{11}~\msunh$ are masked. Curves from red to blue represent increasing galaxy number density, from the sparsest sample (galaxies with $ M_* \geq 10^{11}~\msunh$) to the densest sample (galaxies with $ M_* \geq 10^{7}~\msunh$). The innermost point is plotted at the bin center and does not indicate sub-pixel resolution (see Fig. \ref{fig:Fstack_mask}).}
    \label{fig:Fstack_complete}
\end{figure}

\subsection{Filament Stacking}

The radial density profile obtained through filament stacking characterizes the density distribution perpendicular to the filament spine. Figure \ref{fig:Fstack_mask} shows this profile for different halo masking thresholds. These profiles are strongly concentrated toward the spine, in contrast to the more extended distributions typically observed in pair-stacked filaments. Within a $10~\kpch$ radius, the stacked column density can be up to a factor of six higher than at $100~\kpch$. As a result, the effectiveness of filament stacking methods heavily depends on the observational resolution, since only sufficiently high resolution can resolve the central density peak.

The stacking signal in filament stacking is also affected by halo masking. As shown in Figure~\ref{fig:Fstack_mask}, the unmasked signal at a radius of $d_{\rm fil}=10~\kpch$ is $1.71\times10^{18}~\mathrm{cm^{-2}}$. Masking \hi~in the most massive halos with $M_h \geq 10^{14}~\msunh$ does not significantly alter the stacked density, since these halos occupy a small volume despite their high column density. In contrast, when \hi~is masked in halos with $M_h \geq 10^{11}~\msunh$, the average column density within $10~\kpch$ decreases to $8\%$ ($1.32\times10^{17}~\mathrm{cm^{-2}}$) of the unmasked value, while the density at $100~\kpch$ decreases to $15\%$. \hi~in halos with $M_h = 10^{11\text{--}14}~\msunh$ is highly concentrated at the centers of the filament spines, so removing these halos results in a smoother stacked density profile. As for halos with $M_h = 10^{9\text{--}11}~\msunh$, the \hi~column density at the center of the spine is comparable to that in the intergalactic medium. These halos are more evenly distributed perpendicular to the filament spine and contribute significantly to the outer regions of the stacked filaments. Meanwhile, the amount of \hi~in halos with $M_h < 10^{9}~\msunh$ is negligible, likely due to their proximity to the resolution limit of the simulation.

The effectiveness of filament stacking largely depends on how accurately filaments are identified, and improvements in filament identification directly enhance the amplitude of the stacked signal. A straightforward way to improve filament identification is to increase the galaxy number density of the catalog by lowering the stellar mass threshold. In Figure \ref{fig:Fstack_complete}, we show how galaxy number density affects the density profile of stacked filaments. Identifying filaments using only the brightest galaxies with $M_* \geq 10^{11}~\msunh$ ($n_{gal}=5.7\times10^{-4} \mathrm{Mpc^{-3}}h^{3}$) often leads to ineffective identification. In these cases, the poorly identified cosmic web appears to connect galaxy clusters without genuine filaments, and the \hi~profile perpendicular to filament spines appears nearly uniform. Increasing the galaxy number density by lowering the minimum stellar mass threshold enables more accurate filament identification by better tracing the underlying matter distribution. This improves the localization of filament spines and allows the detection of smaller filaments that are missed in sparse catalogs. Avoiding false identifications by using catalogs with higher number density leads to a higher column density concentrated around filament spines in the stacked signal. Specifically, within $10~\kpch$ of the spine, higher galaxy number density results in an enhances column density due to precise spine localization. However, at larger radii of approximately $100 ~\kpch$, density decreases with increasing number density for $M_* < 10^{9}~\msunh$ ($n_{gal}=3.5\times10^{-2} \mathrm{Mpc^{-3}}h^{3}$). This behavior arises because improved filament identification increases the accuracy of spine localization, reducing random offsets in the stacking and producing a sharper profile, and is also influenced by the inclusion of smaller and more diffuse filaments. Beyond $1~\mpch$ from the spine, higher number density also leads to a lower stacked column density, since this region contains more voids and fewer previously unidentified filaments.

\subsection{Comparison of two methods}

\begin{figure}
    \centering
    \includegraphics[width=\columnwidth]{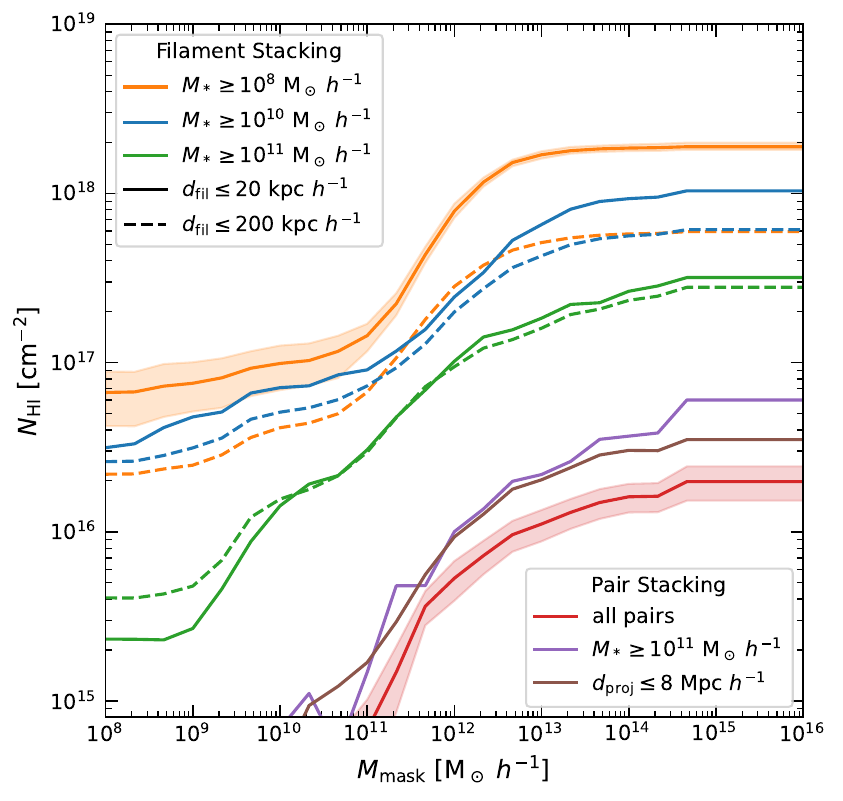}
    \caption{Comparison of two stacking methods. The orange, blue, and green curves represent the results of filament stacking, obtained with galaxy catalogues of different number densities (set by stellar mass thresholds). For these curves, the solid lines show the peak column density within $20~\kpch$ of the filament spine, while the dashed lines show the mean density within $200~\kpch$. The red, purple and brown curves represent the filament signal obtained from pair stacking. The red curve is obtained using all pairs, the purple curve is obtain using pairs connecting the most massive galaxies of $ M_* \geq 10^{11}~\msunh$, and brown curve is obtained using the shortest pairs.}
    \label{fig:Compare}
\end{figure}

Both stacking methods produce \hi~column density, expressed in the same units, and are therefore directly comparable. We present the stacked results in Figure \ref{fig:Compare}. The filament stacking signal is typically 1--2 orders of magnitude higher than that from pair stacking across the full range of masking thresholds. Both techniques are affected by contamination from \hi~in halos, yet the signal from pair stacking diminishes more rapidly as masking becomes more stringent. The effectiveness of filament stacking depends on accurate filament identification, and higher galaxy number density in the catalogue can enhance its performance. For pair stacking, restricting analysis to the most massive or shortest pairs can amplify the filament signal, but this enhancement is largely dominated by \hi~in halos with $ M_h \geq 10^{12}~\msunh$, rather than originating from the diffuse IGM that we aim to probe. The dependence on spatial resolution differs between the two stacking methods. Filament stacking produces a density profile that peaks along the filament spine, and higher resolution increases the stacked column density. In contrast, pair stacking is less sensitive to spatial resolution, as the filament signal in the stacked residual map is relatively smoothly distributed around the center of the pair.

Given current observational capabilities, we now provide a quantitative comparison between filament stacking and pair stacking. State-of-the-art galaxy surveys such as DESI BGS \citep{2023AJ....165..253H} are expected to achieve a comoving galaxy number density of $n_{gal} \geq 10^{-2}~\mathrm{{Mpc}^{-3}}h^3$ at low redshifts, corresponding to a minimum stellar mass of $M_* \sim 10^{9\text{--}10}~\msunh$. Meanwhile, single-dish radio telescopes such as FAST can achieve a spatial resolution of $\sim 26~\kpch$ at these redshifts. In Figure \ref{fig:Compare}, these levels of galaxy number density and spatial resolution correspond to the blue solid curve for filament stacking and the red solid curve for pair stacking. Without masking, the filament stacking signal ($1.04\times10^{18}~\mathrm{cm^{-2}}$) is $\sim 53$ times higher than that from pair stacking ($1.98\times10^{16}~\mathrm{cm^{-2}}$). When MW-size halos ($M_h \geq 10^{12}~\msunh$) are masked, the filament stacking signal ($2.44\times10^{17}~\mathrm{cm^{-2}}$) remains $\sim 46$ times higher than that from pair stacking ($5.31\times10^{15}~\mathrm{cm^{-2}}$). When lower-mass haloes are masked, the difference increases to $\sim 2$ orders of magnitude. Overall, the filament stacking signal exceeds that of pair stacking by $\sim 1.5$--$2$ orders of magnitude across all masking strengths. This indicates that directly identifying filaments using current galaxy catalogues is more effective than relying on the assumption that filaments are traced by galaxy pairs, and the resolution of current single-dish telescopes is sufficient to exploit the advantages of filament stacking.

\section{Discussions}
\label{sec:discussions}

\begin{figure}
    \includegraphics[width=\columnwidth]{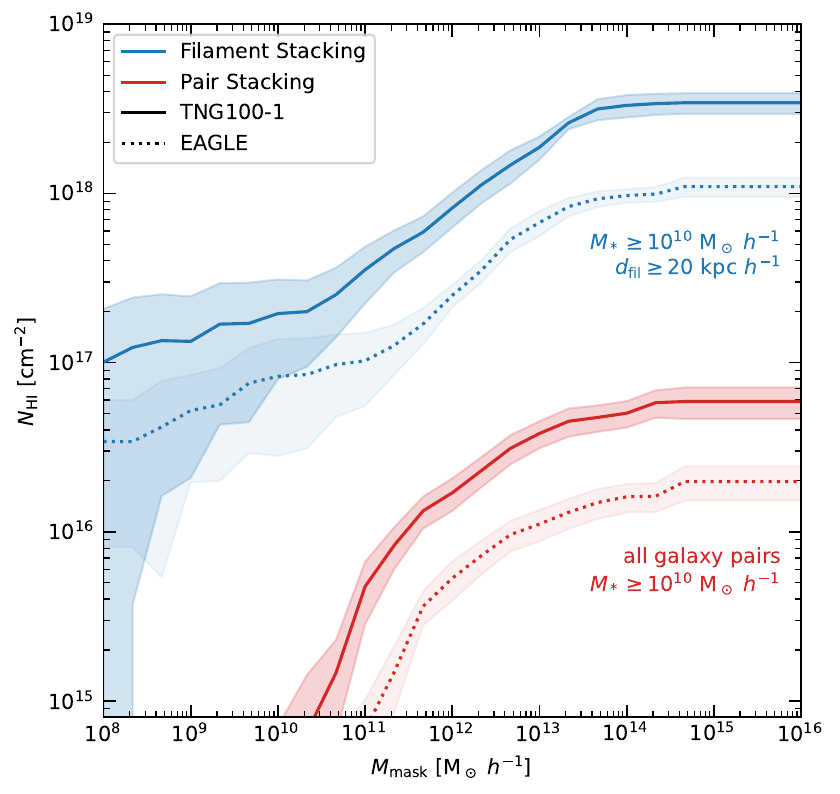}
    \caption{Comparison of two stacking methods in EAGLE and IllustrisTNG simulations. The blue curve show the mean \hi~density within $20~\kpch$ around the filaments identified using galaxies with $ M_* \geq 10^{10}~\msunh$. The red curve is the filament signal in all pairs with galaxies of $ M_* \geq 10^{10}~\msunh$. The solid curves represents results from the TNG100-1 simulation, while the dashed curves represent results from the EAGLE simulation.}
    \label{fig:TNG}
\end{figure}

When predicting \hi~content in simulations, caution is required due to the simplified treatment of physical processes compared to reality. Figure \ref{fig:TNG} illustrates the predicted \hi~column density using both stacking methods applied in the EAGLE and IllustrisTNG simulations. Both simulations rely on post-processed data that are calibrated to reproduce observations of bright galaxies, leaving significant uncertainty regarding \hi~in low mass structures and the IGM. The models predict different results, with IllustrisTNG presenting a \hi~column density approximately half an order of magnitude greater than EAGLE's prediction. Nevertheless, the relative differences between filament stacking and pair stacking methods are consistent across the two simulations. This suggests that while the absolute \hi~density depends on the adopted subgrid physics, filament stacking is more effective by approximately $1.5$--$2$ orders of magnitude compared to pair stacking.

We adopted several simplifying assumptions in our analysis:

1. We conduct our analysis in real space, whereas observations occur in redshift space, which experiences redshift space distortion (RSD). This distortion affects both galaxies and \hi~in the large-scale structure, as well as the galaxy pairs and galaxy-traced cosmic filaments used for stacking. RSD may degrade the accuracy of filament identification and therefore requires careful consideration in observational applications. Nevertheless, we expect that filament identification based on projected galaxy distributions is only mildly affected by RSD.

2. We assume that the observed 21 cm brightness temperature is proportional to the \hi~column density under the optically thin approximation. This assumption is valid for the vast majority of galactic and intergalactic \hi~\citep[e.g.,][]{2012RPPh...75h6901P}. Optical depth effects may become non-negligible in regions of high column density and low spin temperature within galaxies \citep[e.g.,][]{2009ApJ...695..937B, 2017MNRAS.468.4030S}. In such cases, self-absorption can reduce the emergent 21 cm emission relative to the optically thin prediction. However, these regions contribute only a minor fraction of the total \hi~mass budget in the local Universe, and thus the impact on our results is expected to be minimal.

3. Additional observational effects, such as finite beam size and instrumental noise, may further reduce the detectability of the signal. A more realistic assessment requires mock observations that incorporate these effects \citep[e.g.,][]{2025ApJ...984..177L}.

\section{Conclusions}
\label{sec:conclusions}

This study compares two stacking techniques applied to the EAGLE and IllustrisTNG simulations. The pair stacking method is based on the assumption that cosmic filaments connect massive halos, and includes an additional subtraction of a spherical profile to isolate the filament contribution. Meanwhile, filament stacking utilizes the cosmic web identified by galaxy distribution and is heavily dependent on effective filament identification. Both techniques share the objective of enhancing the detection of faint filaments by increasing the S/N through stacking.

We summarize our conclusions regarding these two stacking methods as follows.

\begin{itemize}
    \item[$\bullet$] Both stacking processes are affected by contamination from \hi~in halos. When the majority of halos are masked, filament stacking retains $\sim 10\%$ of its original unmasked value. However, the filament signal in pair stacking is strongly suppressed and becomes statistically insignificant.
    \item[$\bullet$] Filament stacking produces a density profile that peaks at the filament spine's center, whereas in pair stacking, the filament signal is more smoothly distributed. Consequently, the signal of filament stacking are more sensitive to the spatial resolution of the observations.
    \item[$\bullet$] The selection of pairs can enhance the filament signal in pair stacking, particularly when selecting the most massive or shortest pairs. However, once the contribution from halos is excluded, this enhancement becomes insignificant.    \item[$\bullet$] The results of filament stacking are significantly influenced by the galaxy number density of the catalog (set by the stellar mass threshold). Catalogues with higher number density allow for the identification of more filaments and improve the accuracy of filament spine localization.
\end{itemize}

The two stacking methods are based on distinct principles. The pair stacking approach is relatively straightforward to implement and is less sensitive to pair selection. However, it is strongly affected by \hi~contamination from massive halos and struggles to isolate \hi~within smaller halos or in cosmic filaments' diffuse phases. On the other hand, filament stacking, despite its dependence on the completeness of the galaxy catalogue, is more effective overall. Modern galaxy surveys enables more accurate cosmic filament identification, which supports \hi~detection through filament stacking, particularly after masking the most massive halos. Furthermore, higher-resolution observations with radio interferometers are expected to better resolve \hi~filaments, as the \hi~is concentrated along the filament spine. Overall, with current galaxy catalogues, filament stacking is more effective at low redshift.

\begin{acknowledgments}
We thank the anonymous referee for their constructive comments, which helped improve this manuscript. This work is supported by the National Key R\&D Program of China (2022YFA1602901), the NSFC grant (Nos. 12588202, 11873051, 12125302, 11903043) and CAS Project for Young Scientists in Basic Research Grant (No. YSBR-062).
\end{acknowledgments}

\vspace{5mm}
\software{Astropy \citep{2013A&A...558A..33A,2018AJ....156..123A}, 
          \disperse~\citep{2011MNRAS.414..350S,2011MNRAS.414..384S},
          Matplotlib \citep{Hunter2007},
          Numpy \citep{Harris2020},
          Scipy \citep{Virtanen2020}.
          }

\appendix
\section{Masking galaxies}
\label{app:mask_gal}

\begin{figure}
    \centering

    \begin{minipage}{0.495\textwidth}
        \centering
        \includegraphics[width=\textwidth]{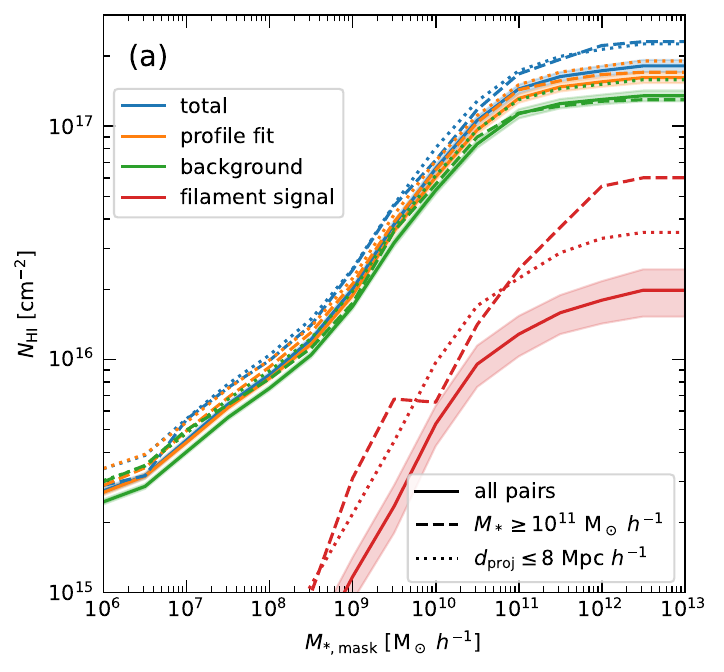}
    \end{minipage}
    \begin{minipage}{0.48\textwidth}
        \centering
        \includegraphics[width=\textwidth]{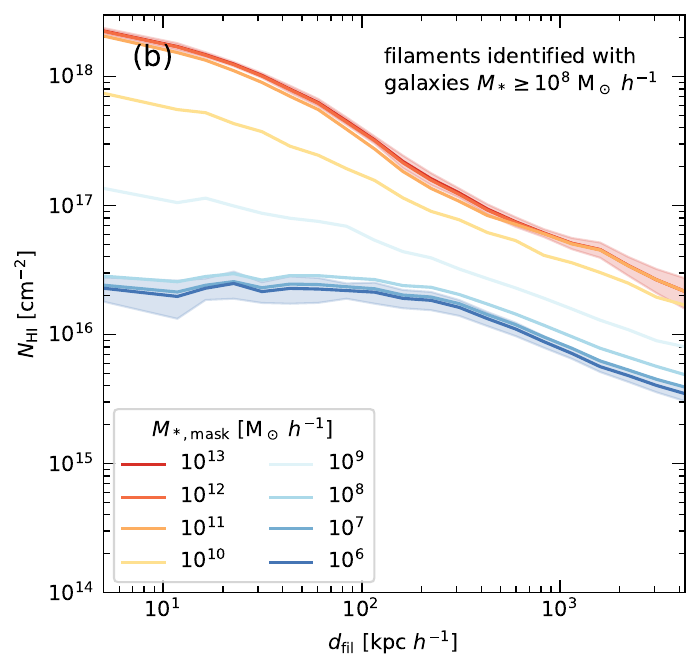}
    \end{minipage}

    \vspace{0.3cm}

    \hspace{0.24cm}
    \begin{minipage}{0.48\textwidth}
        \centering
        \includegraphics[width=\textwidth]{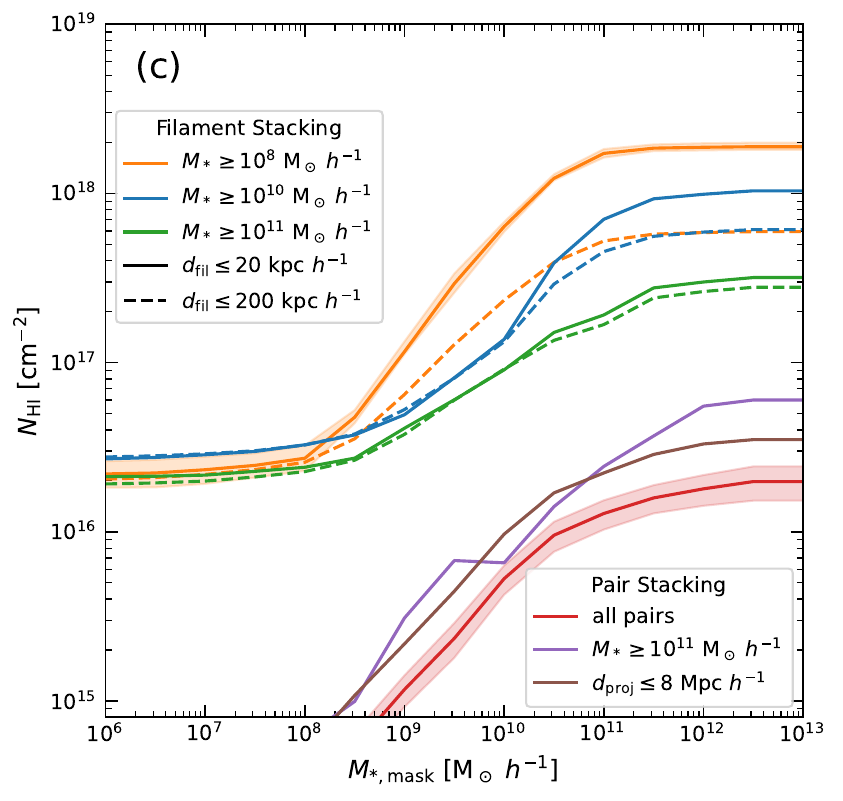}
    \end{minipage}
    \hspace{0.19cm}
    \begin{minipage}{0.48\textwidth}
        \centering
        \includegraphics[width=\textwidth]{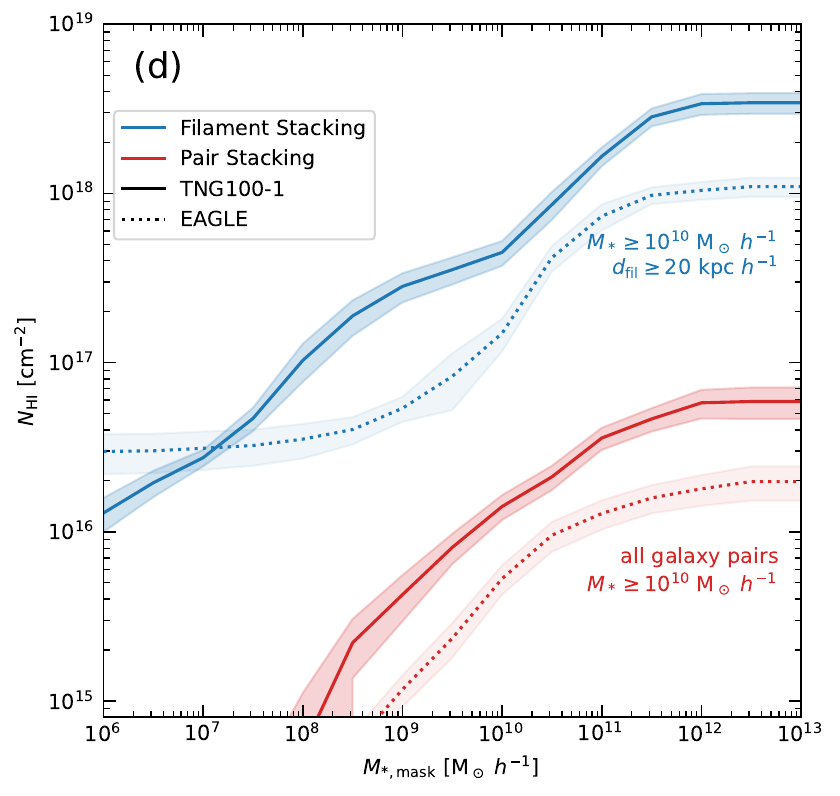}
    \end{minipage}

    \caption{
    Results of masking galaxies (subhalos) above a stellar mass threshold $M_{\rm *, mask}$.
    (a) Pair stacking results, analogous to Figure \ref{fig:Pstack}.
    (b) Filament stacking results, analogous to Figure \ref{fig:Fstack_mask}.
    (c) Comparison of two stacking methods, analogous to Figure \ref{fig:Compare}.
    (d) Comparison between IllustrisTNG and EAGLE simulations, analogous to Figure \ref{fig:TNG}.
    }
    \label{fig:App}
\end{figure}

Here we present an alternative masking scheme in which galaxies (subhalos) above a given stellar mass threshold are removed. We mask regions within twice the gas half-mass radius of the selected subhalos and then apply the same averaging and stacking procedures as in the main text.

The results are shown in Figure~\ref{fig:App}. We find that \hi~associated with galaxies contributes a substantial fraction of the stacked filament signal. For pair stacking (panel~a), masking galaxies with $M_* \geq 10^{10}~\msunh$ reduces the filament signal to $27\%$ ($5.28\times10^{15}~\mathrm{cm^{-2}}$), while masking galaxies with $M_* \geq 10^{9}~\msunh$ further reduces it to $6\%$ ($1.17\times10^{15}~\mathrm{cm^{-2}}$); applying more stringent masking lowers the signal to below $10^{15}~\mathrm{cm^{-2}}$. The enhancement of the filament signal from pair selection becomes less significant once galaxies are masked, consistent with the trend observed in halo masking.

For filament stacking (panel~b), the mean \hi~column density at the filament spine decreases to $32\%$ ($5.54\times10^{17}~\mathrm{cm^{-2}}$) and $6\%$ ($1.05\times10^{17}~\mathrm{cm^{-2}}$) when masking galaxies with $M_* \geq 10^{10}~\msunh$ and $M_* \geq 10^{9}~\msunh$, respectively. Compared to halo masking, galaxy masking has a stronger effect on the central spine region but a weaker impact on the outer filament regions.

Comparing the two methods (panel~c), we find that filament stacking yields a signal that is $1$--$2$ orders of magnitude higher than that of pair stacking, and similar trends are observed in the IllustrisTNG simulation (panel~d). The relative strengths of filament stacking and pair stacking are consistent with those found in halo masking.


\bibliography{references}{}
\bibliographystyle{aasjournalv7}

\end{document}